\documentclass{elsart}

\usepackage{epsfig}

\begin{document}
  
  \begin{frontmatter}

    \begin{flushleft}
      \footnotesize{\tt WUE-ITP-2005-014 }
    \end{flushleft}
  
    \vspace{4cm}

    \title {Automatized analytic continuation of Mellin-Barnes
    integrals}

    \author[MC]{M. Czakon}

    \corauth[MC]{\tt e-mail: mczakon@physik.uni-wuerzburg.de}

    \address{Institut f\"ur Theoretische Physik und Astrophysik,
    Universit\"at W\"urzburg, \\ Am Hubland, D-97074 W\"urzburg,
    Germany}

    \address{Department of Field Theory and Particle Physics,
      Institute of Physics, \\ University of Silesia, Uniwersytecka 4,
      PL-40007 Katowice, Poland}
  
    \begin{abstract}

      I describe a package written in {\tt MATHEMATICA} that
      automatizes typical operations performed during evaluation of
      Feynman graphs with Mellin-Barnes (MB) techniques. The main
      procedure allows to analytically continue a MB integral in a
      given parameter without any intervention from the user and thus
      to resolve the singularity structure in this parameter. The
      package can also perform numerical integrations at specified
      kinematic points, as long as the integrands have satisfactory
      convergence properties. I demonstrate that, at least in the case
      of massive graphs in the physical region, the convergence may
      turn out to be poor, making na\"ive numerical integration of MB
      integrals unusable. I present possible solutions to this
      problem, but argue that full automatization in such cases may
      not be achievable.
  
    \end{abstract}

  \end{frontmatter}

  \newpage

  \section*{PROGRAM SUMMARY}

   {\it Title of program:} {\tt MB}

   {\it Version:} {\tt 1.1}

   {\it Catalogue identifier:}

   {\it Program obtainable from:} {\tt
   http://theorie.physik.uni-wuerzburg.de/\~{}mczakon}

   {\it Computers:} All

   {\it Operating systems:} All

   {\it Programming language used:} {\tt MATHEMATICA}, {\tt Fortran
   77} for numerical evaluation

   {\it Memory required to execute with typical data:} Sufficient for
   a typical installation of {\tt MATHEMATICA}.

   {\it No. of bytes in distributed program, including test data:}
   337900

   {\it Distribution format:} {\tt ASCII}

   {\it Libraries used:} {\tt CUBA} \cite{cuba} for numerical
   evaluation of multidimensional integrals and {\tt CERNlib}
   \cite{cernlib} for the implementation of $\Gamma$ and $\psi$
   functions in {\tt Fortran}.

   {\it Keywords:} Mellin-Barnes integrals, analytic continuation,
   numerical evaluation, Feynman integrals.

   {\it Nature of physical problem:} Analytic continuation of
   Mellin-Barnes integrals in a parameter and subsequent numerical
   evaluation . This is necessary for evaluation of Feynman integrals
   from Mellin-Barnes representations.

   {\it Method of solution:} Recursive accumulation of residue terms
   occurring when singularities cross integration contours. Numerical
   integration of multidimensional integrals with the help of the {\tt
   CUBA} library.

   {\it Restrictions on the complexity of the problem:} Limited by the
   size of the available storage space.

   {\it Typical running time:} Depending on the problem. Usually
   seconds for moderate dimensionality integrals.

  \newpage

  \section{Introduction}

  The synergy between experiment and theory in the area of elementary
  particle physics is constantly driving perturbative calculations to
  higher and higher orders. This is particularly true close to the
  beginning of the Large Hadron Collider's operation. Therefore,
  recent years have seen the emergence of several powerful methods of
  evaluation of subsequent terms of the perturbative expansion. As far
  as multiloop Feynman integrals are concerned, the method of
  differential equations Ref.~\cite{Kotikov:1991pm,Remiddi:1997ny} and
  Mellin-Barnes  integral representations
  Ref.~\cite{Smirnov:1999gc,Tausk:1999vh} have proved to be the most
  successful. Some complicated problems turned out to even require a
  mixed approach, as advocated, for example, in the case of Bhabha
  scattering in Ref.~\cite{proceedings}. In parallel to analytical
  approaches, new numeric techniques have been devised, among which
  the sector decomposition method Ref.~\cite{Binoth:2000ps} occupies a
  prominent place. Very recently in Ref.~\cite{Anastasiou:2005cb}, the
  role of Mellin-Barnes integral representations as sources of numeric
  approximations in the physical region has also been stressed.

  In the present work, I will concentrate on Mellin-Barnes integral
  representations. There are two advantages of this approach. First,
  it allows for systematic extraction of singularities. Second, the
  dimensionality of the representation is not directly connected to
  the number of lines in the graph and therefore, one often arrives at
  integrals of low dimensionality even for complicated graphs. The
  calculation of a Feynman integral proceeds in this method in three
  steps. At first, one derives a representation, then performs the
  analytic continuation in $\epsilon$, where $d=4-2\epsilon$ is the
  dimension of spacetime, and finally evaluates the resulting
  integrals. The first step above can be performed in several
  different ways, aiming at the simplest possible representation. The
  various possibilities are described in Ref.~\cite{Smirnov:2004ym}. A
  general algorithm is here only interesting in the case of subsequent
  numeric integration, see Ref.~\cite{Anastasiou:2005cb}. The third
  step cannot be generalized apart from numerical integration, even
  though some classes of problems can be solved algorithmically, {\it
  e.g.} by reduction to nested sums, see Ref.~\cite{Moch:2001zr}. It
  is only in the second step, the analytic continuation, that one can
  provide an algorithmic solution that would be satisfactory for both
  analytic and numeric evaluation. This solution is provided by the
  {\tt MATHEMATICA} package {\tt MB} introduced in the present work.

  Numeric evaluation of MB integrals has already been mentioned more
  than once above. Whether just for testing or for the actual
  calculation, automatization of this step is of value by itself. The
  package MB can perform the necessary integration by means of
  FORTRAN, the CUBA library \cite{cuba} of integration routines, and
  the CERN library implementation of gamma and psi functions
  \cite{cernlib}. Since the integrals are infinite range and
  multidimensional, their feasibility depends strongly on their
  convergence. In all tested examples, where invariants are in the
  Euclidean range, the behaviour is exponential and therefore poses no
  problems. In \cite{Anastasiou:2005cb}, physical kinematics have also
  been considered, but the presented examples were restricted to
  massless graphs exclusively. Here, I notice that massive graphs have
  worse properties. In fact, I give examples of integrals, which are
  not even absolutely integrable, and the integral is similar to the
  Fourier transform of the inverse square root. Such cases can still
  be treated, but some initial analysis is necessary and it is
  difficult to see how it could be automatized. Moreover, the
  techniques will rapidly become inefficient for higher dimensional
  integrals.

  The paper is organized as follows. In the next section, I define the
  main concepts and present the algorithm for analytic
  continuation. Subsequently, I describe the package starting with the
  user interface, low level routines, examples, numerical integration
  routines and some additional tools. Finally, I briefly summarize and
  conclude the paper.

  \section{Analytic continuation of Mellin-Barnes integrals}

  At the core of the Mellin-Barnes method lies the following
  representation

  \begin{equation}
    \frac{1}{(A+B)^\nu} = \frac{1}{\Gamma(\nu)}\frac{1}{2\pi
    i}\int_{-i \infty}^{i \infty} dz \frac{A^z}{B^{\nu+z}}
    \Gamma(-z)\Gamma(\nu+z),
    \label{mb}
  \end{equation}

  where the contour is chosen in such a way, that the poles of the
  $\Gamma$ function with $+z$ are separated from the poles of the
  $\Gamma$ function with $-z$.

  This representation can be used in Feynman integral computations in
  several ways. The easiest is to turn massive propagators into
  massless and integrate the massless integral, if a formula for
  general powers of propagators exists. In more complicated cases, one
  can use some parametric representation of the Feynman integral,
  which is usually an integral of a product of polynomials raised to
  some powers, and split the polynomials into pieces that are then
  integrable by some generalization of the Euler formula

  \begin{equation}
    \int_0^1 dx \; x^{\alpha-1}(1-x)^{\beta-1} =
    \frac{\Gamma(\alpha)\Gamma(\beta)}{\Gamma(\alpha+\beta)}.
  \end{equation}

  An extensive discussion of the methods with examples can be found in
  Ref.~\cite{Smirnov:2004ym}. Irrespective of the method, however, the
  expression for any Feynman integral assumes the form

  \begin{eqnarray}
    \!\!\!\!\!\!\!\!\!\!\!\!  \frac{1}{(2\pi i)^n}\int_{-i \infty}^{i
    \infty} \dots \int_{-i \infty}^{i \infty} \Pi_i dz_i \;
    f(z_1,\dots,z_n,s_1,\dots,s_p,a_1,\dots,a_q,\epsilon)  \frac{\Pi_j
    \Gamma(A_j+V_j+c_j \epsilon)}{\Pi_k \Gamma(B_k+W_k+d_k \epsilon)},
    \nonumber\\ \label{definition}
  \end{eqnarray}
  where $s_i$ are some kinematic parameters and masses; $a_i$ are the
  powers of the propagators; $A_i$, $B_i$ are linear combinations of
  the $a_i$; $V_i$, $W_i$ are linear combinations of $z_i$; and $c_i$,
  $d_i$ are some numbers. The function $f$ is analytic, in practice a
  product of powers of the $s_i$, with exponents being linear
  combinations of the remaining parameters.

  Because of the assumptions inherent in Eq.~(\ref{mb}), the above
  equation is well defined and corresponds to the original Feynman
  integral, if the real parts of all of the $\Gamma$ functions have
  positive arguments. If these conditions cannot be satisfied with
  $\epsilon = 0$, then the integral may develop divergences and
  analytic continuation to $0$ is necessary to make an expansion in
  $\epsilon$.

  The purpose of the presented package is to perform the analytic
  continuation of Eq.~(\ref{definition}) in $\epsilon$ to some chosen
  value $\epsilon_0$. The algorithm requires to generalize
  Eq.~(\ref{definition}) to allow for $\psi$ functions in the
  fraction, with $\psi(z)  = d \log \Gamma(z) /d z$ and $\psi^{(n)}(z)
  = d^{n} \psi(z)/dz^{n}$, with the same structure of arguments as
  those of the  $\Gamma$ functions.

  \subsection{The algorithm}

  \label{algorithm}

  There are two known ways to perform the analytic continuation. The
  first, introduced in Ref.~\cite{Smirnov:1999gc} consists in
  deforming the integration contours and then shifting them past the
  poles of the $\Gamma$ functions, which results in residue
  integrals. It is not clear how to make this method algorithmic,
  although some attempts in the specific case of massless on-shell
  double boxes have been undertaken in \cite{Smirnov:1999wz}.

  The second method, introduced in Ref.~\cite{Tausk:1999vh} assumes
  fixed contours parallel to the imaginary axis, and the analytic
  continuation consists in accounting for pole crossings past the
  contours. As described in Ref.~\cite{Tausk:1999vh}, this method is
  an algorithm. I make one modification with respect to the original,
  namely I assume that the contours are such that no two contours can
  be crossed simultaneously. This assumption can always be satisfied
  by infinitesimal shifts of one of the concerned contours.

\begin{figure}
    \begin{center}
      \fbox{\epsfig{file=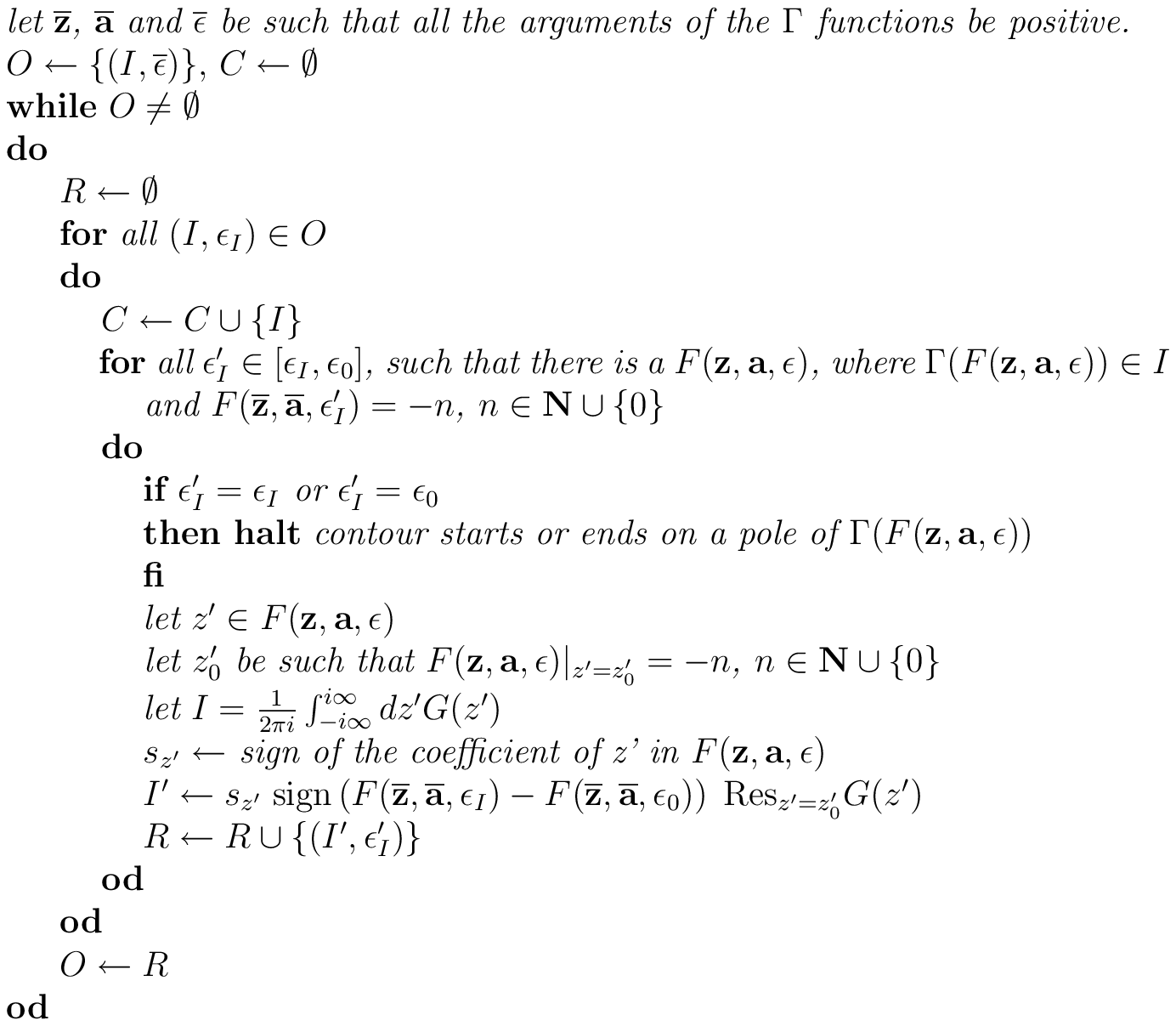}}
    \end{center}
    \caption{\label{algo} \it Analytic continuation algorithm.}
  \end{figure}

  It should be clear from the above considerations, that the imaginary
  parts of the involved variables do not play any role. It is
  therefore assumed that $z_i$, $a_i$ and $\epsilon$ are real.  With
  ${\bf z} = (z_1, \dots, z_n)$, ${\bf a} = (a_1, \dots, a_q)$, and
  $(I,\epsilon_I)$ some MB integral with fixed contours and the value
  of $\epsilon$ fixed at $\epsilon_I$, the algorithm can be formalized
  as in Fig.~\ref{algo}.

  The algorithm has been written for $\Gamma$ functions, but one
  should add $\psi$ functions, wherever $\Gamma$ functions occur.
  Upon termination, the set $C$ contains all the integrals following
  from the analytic continuation. It should be clear that it is the
  ``if'' clause that does not allow for crossings of two different
  contours at a time.

  A comment about the choice of the contours is in order. Even though
  all the choices are equivalent, one would like to have the smallest
  possible number of contributions. An improvement implemented in the
  package is to first gather all the residue points, and then try to
  add additional constraints on the contours such that these residues
  would not occur. If some subset of these constraints can be
  satisfied, then the number of residues will be reduced. This is not
  an algorithm that leads to an absolute minimum of the number of
  residues, it gives, however, at least some reduction of the number
  of contributions.

  Finally, one should notice that the technique of
  Ref.~\cite{Tausk:1999vh} has been similarly formalized in
  Ref.~\cite{Anastasiou:2005cb}.

  \section{The package}

  \subsection{User interface}

  The main routine performing the analytic continuation is

  \fbox{\tt MBcontinue[integrand, limit, \{fixedVarRules,
  intVarRules\}, options]}

  where the input arguments are

  \begin{itemize}

    \item {\tt integrand:} any object accepted by {\tt
    MATHEMATICA}. Notice that the singularities are determined by
    analyzing $\Gamma$ and $\psi$ functions only.

    \item {\tt limit:} a rule, {\tt x -> x0}, which specifies at the
    same time the variable, {\tt x}, in which the analytic
    continuation is performed and the point, {\tt x0}, which the user
    wants to reach.

    \item {\tt fixedVarRules:} a list of rules giving the values of
    the real parts of the variables, which are not integrated over. In
    particular, it must contain the starting value of the variable, in
    which the analytic continuation is performed.

    \item {\tt intVarRules:} a list of rules giving the real parts of
    the integration variables.

    \item {\tt options:}

      \begin{itemize}

	\item {\tt Level:} an integer specifying the level at which
	the recursive analytic continuation will be stopped. By
	default, it is set to infinity.

	\item {\tt Skeleton:} a boolean value. If {\tt True}, the
	residues will be identified, but not calculated. This is
	achieved by replacing all $\Gamma$ and $\psi$ functions by a
	dummy function {\tt MBgam}. The purpose of this option is to
	quickly determine the total number of integrals. By default
	this option is set to {\tt False}.

	\item {\tt Residues:} a boolean value. If {\tt True}, the
	output will also contain the list of Residue points besides
	the actual values of the residues. This is mainly for internal
	use and is set by default to {\tt False}.

	\item {\tt Verbose:} a boolean value. If {\tt True}, the level
	is printed as well as the position on the list of the
	currently continued integral and the residue points together
	with the signs of the residues. This option is switched on by
	default.

      \end{itemize}

  \end{itemize}

  The output is a nested list obtained by replacing, at every level,
  the integral to be continued by its residues and the original
  integral at the limit. The elements are

  {\tt MBint[integrand, \{fixedVarRules, intVarRules\}]}
    
  objects, where the integrand can be expanded around the limit, which
  is placed on the {\tt fixedVarRules} list. If the user specified a
  finite level, then there might also occur
  
  {\tt MBitc[integrand, limit, \{fixedVarRules, intVarRules\},
      Options]}
  
  objects, where ``itc'' stands for ``integral to continue''. These
  are not yet regular at the limit and require further recursive
  analytic continuation. Furthermore, if the user set the {\tt
  Residues} option to {\tt True}, there will also be a list of

  {\tt MBres[sign, var, val]},

  objects, which signal that there was a residue taken in the
  variable, {\tt var}, at the value, {\tt val}, with {\tt sign}.

  Restricted input checking has been implemented, and as long as the
  input is syntactically correct, the only error that may occur is
  (see Section~\ref{algorithm} for further details)

  {\tt contour starts and/or ends on a pole of Gamma[z]}

  In this case the procedure stops and gives an inequality for an
  integration variable that is sufficient to remove the problem.

  The integration contours are found with

  \fbox{\tt MBoptimizedRules[integrand, limit, constraints, fixedVars,
  options]}

  For a description of the {\tt integrand} and {\tt limit} see {\tt
  MBcontinue}. The remaining input parameters are as follows

  \begin{itemize}

    \item {\tt constraints:} a list of additional constraints
    (inequalities) specified by the user. This should usually be left
    empty, but might be used for experimentation in order to search
    for contours that might possibly give less residues.

    \item {\tt fixedVars:} a list of variables, which should be
    considered fixed during analytic continuation. The integration
    variables are determined automatically from the arguments of the
    $\Gamma$ and $\psi$ functions.

    \item {\tt options:}

      \begin{itemize}

	\item {\tt Level:} specifies the level up to which
	optimization of the contours will be performed. This option
	should only be used for very large calculations. Since
	in this case, the contours  are only partially tested, the
	user will have to correct them himself, if poles lying on a
	contour are encountered. In practice, independent, small
	shifts should be sufficient for this purpose.

      \end{itemize}

  \end{itemize}

  The output matches precisely the form needed in the input of {\tt
  MBcontinue}, {\it i.e.}

  {\tt \{fixedVarRules, intVarRules\}}

  Notice that this procedure not only reduces the number of residues,
  but also generates such contours that, during analytic continuation,
  no contours will start or end on a pole.

  During the determination of the real parts, warning messages are
  generated. These can be ignored apart from the case when there is a
  single message

  {\tt no rules could be found to regulate this integral}

  and the output is an empty list. In this case, the integral cannot
  be regulated and the user has to provide another one, {\it e.g.} by
  introducing a further regulator parameter, for example a propagator
  power, and performing two subsequent analytic continuations.

  Once the integrals are determined, they can be either {\bf merged},
  {\it i.e.} those that have the same contour will be added by
  linearity; {\bf preselected}, {\it i.e.} those that would vanish in
  a given order of expansion in some parameter are rejected; or {\bf
  expanded}. These tasks are achieved with the following utilities.

  \fbox{\tt MBmerge[integrals]}

  Merges {\tt MBint} objects on the {\tt integrals} list by linearity,
  if they have the same contours. Vanishing integrals are rejected.

  \fbox{\tt MBpreselect[integrals, \{x, x0, n\}]}

  Rejects those {\tt MBint} objects on the {\tt integrals} list that
  would vanish after expansion in the variable {\tt x}, around the
  point {\tt x0}, up to order {\tt n}.

  \fbox{\tt MBexpand[integrals, norm, \{x, x0, n\}]}
  
  Expands {\tt MBint} objects on the {\tt integrals} list around the
  point {\tt x0}, in the variable {\tt x}, up to order {\tt n}. A
  normalization factor, {\tt norm}, is included in every integrand.

  \subsection{Low level routines}

  The routines described in the previous section form the
  interface. It might happen that the user would like to use the low
  level routines, which actually perform the calculation.

  \fbox{\tt MBresidues[integrand, limit, \{fixedVarRules,
  intVarRules\}, options]}

  Performs a single step in the recursive analytic continuation
  algorithm, {\it i.e.} it finds all the residues for a given
  integral, but does not proceed with the analytic continuation of the
  resulting integrals. All the arguments and options are the same as
  in {\tt MBcontinue}, apart from {\tt Level}, which is in this case
  meaningless.

  \fbox{\tt MBrules[integrand, constraints, fixedVars]}

  Finds the real parts of all the fixed and integration variables,
  such that the real parts of the arguments of all the $\Gamma$ and
  $\psi$ functions be positive. The difference to {\tt
  MBoptimizedRules} is that no attempt is made to optimize the number
  of residues or even check whether the contours will not lead to
  problems with {\tt MBcontinue}. To perform these tests, {\tt
  MBoptimizedRules} needs the limit of the continuation, which is left
  unspecified here. This routine is of particular interest, because
  one may use it to write another contour optimization algorithm.

  \fbox{\tt MBrules[integrand, limit, constraints, fixedVars]}

  Same as {\tt MBrules}, but check the contours, so that a complete
  analytic continuation with {\tt MBcontinue} can be performed.

  \subsection{Examples}

  \begin{figure}
    \begin{center}
      \epsfig{file=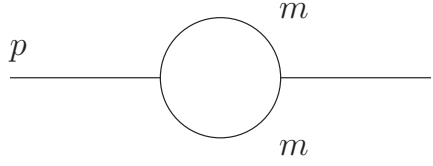, width=6cm}
    \end{center}
    \caption{\label{b0} \it The $B_0(s,ms,ms)$ function, with $s=p^2$
      and $ms=m^2$.}
  \end{figure}
  
  As a first example, I consider the $B_0$ function with two equal
  masses, Fig.~\ref{b0}. After introduction of two MB integrations
  (the integral can be further simplified by the use of the first
  Barnes lemma, see Section~\ref{tools}) and normalization of the
  integration measure with $1/(i \pi^{d/2})$, the expression reads

  {\small
    \begin{verbatim}
In[1]:= int = b0[s, 1+z1, 1+z2]*ms^z1*ms^z2*
        Gamma[-z1]*Gamma[1+z1]*Gamma[-z2]*Gamma[1+z2] /. z1 -> z1-z2

Out[1]:= (m1s^z1*(-s)^(-ep - z1)*Gamma[ep + z1]*Gamma[1 - ep - z2]*
         Gamma[-z2]*Gamma[-z1 + z2]*Gamma[1 - ep - z1 + z2])/ Gamma[2
         - 2*ep - z1]
    \end{verbatim}
  }

  The user must now determine the contours, or more precisely, the
  real parts of the contours.

  {\small
    \begin{verbatim}
In[2]:= rules = MBoptimizedRules[int, ep -> 0, {}, {ep}]

MBrules::norules: no rules could be found to regulate this integral

MBrules::norules: no rules could be found to regulate this integral

Out[2]:= {{ep -> 7/8}, {z1 -> -3/4, z2 -> -1/2}}
    \end{verbatim}
  }

  As explained above, the two warning messages have been generated
  during the determination of the contours, and since some real parts
  have been found, they are harmless.

  The user can now perform the analytic continuation

  {\small
    \begin{verbatim}
In[3]:= cont = MBcontinue[int, ep -> 0, rules]

Level 1

Taking +residue in z1 = -ep

Level 2

Integral {1}

Taking +residue in z2 = -ep

Level 3

Integral {1, 1}

3 integral(s) found

Out[3]:= {{{MBint[(Gamma[1 - ep]*Gamma[ep])/ (m1s^ep*Gamma[2 - ep]),
              {{ep -> 0}, {}}]},  MBint[(Gamma[1 - ep -
              z2]*Gamma[-z2]*Gamma[1 + z2]* Gamma[ep +
              z2])/(m1s^ep*Gamma[2 - ep]),  {{ep -> 0}, {z2 ->
              -1/2}}]},  MBint[(m1s^z1*(-s)^(-ep - z1)*Gamma[ep + z1]*
              Gamma[1 - ep - z2]*Gamma[-z2]*Gamma[-z1 + z2]* Gamma[1 -
              ep - z1 + z2])/Gamma[2 - 2*ep - z1],  {{ep -> 0}, {z1 ->
              -3/4, z2 -> -1/2}}]}
    \end{verbatim}
  }

  At this stage, the user can, for example, expand the integrals to
  determine the divergence

  {\small
    \begin{verbatim}
In[4]:= div = MBexpand[cont, Exp[ep EulerGamma], {ep, 0, -1}]

Out[4]:= {{{MBint[ep^(-1), {{ep -> 0}, {}}]}}}
    \end{verbatim}
  }

  This is the well known value for the $B_0$ function. The integral
  header, {\tt MBint}, is kept, because in general, even the
  divergences may be given by nontrivial MB integrals.
  
  \begin{figure}
    \begin{center}
      \epsfig{file=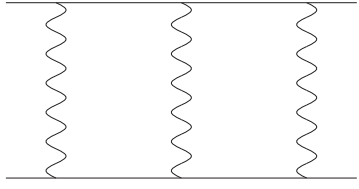, width=5cm}
    \end{center}
    \caption{\label{B1} \it First planar QED box master integral,
      B1. The wavy lines are massless, whereas the continuous are massive
      and on-shell.}
  \end{figure}
  
  \begin{figure}
    \begin{center}
      \epsfig{file=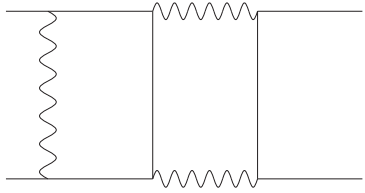, width=5cm}
    \end{center}
    \caption{\label{B2} \it Second planar QED box master integral,
      B2. The notation is the same as in Fig.~\ref{B1}.}
  \end{figure}
  
  Together with the {\tt MB.m} package, two example notebooks are
  provided. The first one, {\tt MBexamples1.nb}, contains massive box
  integrals, in particular, the first and the second planar 7-line QED
  box master integrals, Fig.~\ref{B1} and Fig.~\ref{B2}
  respectively. It is found that in the first case, only 5 integrals
  contribute to the finite part, which is less than has been
  determined in Ref.~\cite{Smirnov:2001cm} by another method of
  analytic continuation. After merging, both integrals have just 4
  contributions. I have checked by numerical integration that the
  results agree with Ref.~\cite{Smirnov:2001cm} and
  Ref.~\cite{Heinrich:2004iq}.

  \begin{figure}
    \begin{center}
      \epsfig{file=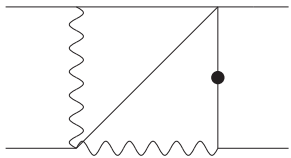, width=4cm}
    \end{center}
    \caption{\label{B5l3md2} \it The B5l3md2 integral. The notation is
      the same as in Fig.~\ref{B1}.}
  \end{figure}
  \begin{figure}
    \begin{center}
      \epsfig{file=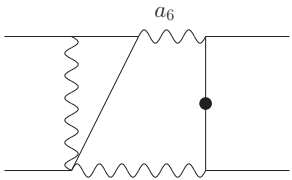, width=4cm}
    \end{center}
    \caption{\label{reg} \it A regularized version of the B5l3md2
      integral.  A finite result can be derived from the general
      representation for B2, when $a_6\rightarrow 0$.}
  \end{figure}
  
  An interesting example is the {\tt B5l3md2} integral,
  Fig.~\ref{B5l3md2}, from Ref.~\cite{Czakon:2004tg}. If one uses the
  general representation from Ref.~\cite{Heinrich:2004iq}, and simply
  sets the powers of the propagators to appropriate values, then the
  integral seems to vanish, due to a $\Gamma$ function in the
  denominator, $1/\Gamma(0) = 0$. To overcome this problem, one keeps
  one of the powers as a parameter, as in Fig.~\ref{reg} and does
  first an analytic continuation in this parameter. In this way, one
  obtains the following MB representation

  \begin{eqnarray*}
    &&\!\!\!\!\!\!\!\!\!\!\!\!\!\!\!\frac{1}{(2 \pi i)^4}
    \frac{1}{\Gamma(-2\epsilon)} \int_{-i \infty}^{i \infty} \int_{-i
      \infty}^{i \infty} \int_{-i \infty}^{i \infty} \int_{-i \infty}^{i
      \infty}\;\; dz_1\; dz_2\; dz_5\; dz_6\; (-s)^{-2 - 2 \epsilon - z_5 -
      z_6} \left(\frac{t}{s}\right)^{z_1} \\ &\times&\frac{ \Gamma(-z_1)
      \Gamma(1 + z_1) \Gamma(-1 - 2 \epsilon - z_2)  \Gamma(1 + z_2)
      \Gamma(-3 - 4 \epsilon - 2 z_1 - z_2 - 2 z_5)}{ \Gamma(-1 - 2 \epsilon
      - z_2 - 2 z_5)  \Gamma(-3 \epsilon - z_5)\Gamma(-3 - 4 \epsilon - 2
      z_1 - z_2 - 2 z_5 - 2 z_6)} \\ &\times&\Gamma(-1 - \epsilon - z_5)
    \Gamma(-\epsilon - z_2 - z_5) \Gamma(-z_5)  \Gamma(2 + \epsilon + z_1
    + z_2 + z_5) \\ &\times&\Gamma(-1 - 2 \epsilon - z_1 - z_5 - z_6)
    \Gamma(-2 - 2 \epsilon - z_1 - z_2 - z_5 - z_6) \Gamma(-z_6) \\
    &\times&\Gamma(2 + 2 \epsilon + z_1 + z_5 + z_6).
  \end{eqnarray*}
  
  The presence of the $1/\Gamma(-2\epsilon)$ factor means that as long
  as we are only interested in the finite part, the integral is just
  threefold. This is, of course, confirmed by explicit continuation as
  can be checked in {\tt MBexamples1.nb}, where three contributions
  are obtained. This result has been numerically checked against the
  one obtained by the sector decomposition method.

  \begin{figure}
    \begin{center}
      \epsfig{file=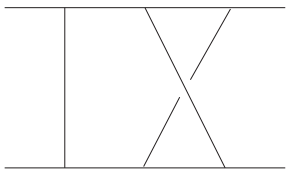, width=4cm}
    \end{center}
    \caption{\label{NP} \it Massless on-shell non-planar double box
      integral, NP.}
  \end{figure}
  \begin{figure}
    \begin{center}
      \epsfig{file=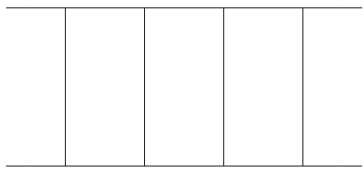, width=5cm}
    \end{center}
    \caption{\label{T} \it Massless on-shell triple box integral, T.}
  \end{figure}
  
  The second notebook, {\tt MBexamples2.nb}, contains two massless
  on-shell box integrals, the two-loop non-planar {\tt NP},
  Fig.~\ref{NP}, and the three-loop planar {\tt T}, Fig.~\ref{T}. In
  the first case, I evaluate the first three poles of the expansion
  and obtain at the symmetric point, $s=-1, t=-1, u=-1$

  \begin{equation}
    \mbox{NP} = \Gamma(3+2\epsilon) \left(
    \frac{7}{4\epsilon^4}-\frac{3}{\epsilon^3}-\frac{1}{\epsilon^2}\left(
    \frac{7}{2} +\frac{47 \pi^2}{24} \right) + \dots\right),
  \end{equation}

  in perfect agreement with Ref.~\cite{Tausk:1999vh}. Similarly, I
  recover the value of the first three poles of the triple-box
  integral Ref.~\cite{Smirnov:2003vi}

  \begin{equation}
    \mbox{T} = -\frac{e^{-3\epsilon \gamma_E} }{s^3(-t)^{1+3\epsilon}}
    \left(\frac{16}{9\epsilon^6} - \frac{5\log(s/t)}{3 \epsilon^5} -
    \frac{3\pi^2}{2\epsilon^4}+\dots\right).
  \end{equation}

  In both cases the lowest order pole was given by one-dimensional
  integrals that could be made with the first Barnes lemma, see
  Section~\ref{tools}.

  \subsection{Numerical integration}

  There are two factors determining the rate of convergence of MB
  integrals Eq.~(\ref{definition}): the behaviour of the product of
  gamma functions for large imaginary arguments and the behaviour of
  the analytic $f$ function.

  In the limit of large imaginary argument, the $\Gamma$ function
  exhibits an oscillatory behaviour, an exponential damping factor and
  a power law. Indeed, for $a,b \in {\rm I\mkern-3mu R}$ and $b \gg 0$ 

  \begin{eqnarray}
    \Gamma(a+i b) &\simeq& \sqrt{2\pi} \; e^{i \frac{\pi}{4} (2a-1)} \; e^{i
    b\; (\log{b} - 1)} \; e^{-\frac{b \pi}{2}} \; b^{a-1/2}, \\ &&
    \nonumber \\
    \Gamma(a-i b) &\simeq& \sqrt{2\pi} \; e^{-i \frac{\pi}{4} (2a-1)} \;
    e^{-i b\; (\log{b} - 1)} \; e^{-\frac{b \pi}{2}} \; b^{a-1/2}.
  \end{eqnarray}

  If we combine different gamma functions the exponential factor might
  in principle disappear, but fortunately in all cases studied it
  did not.

  On the other hand, as explained in Ref.~\cite{Smirnov:2004ym}, the
  $f$ function is usually a product of terms of the form
  
  \begin{equation}
    \label{kinematic}
    (-s)^{-z},
  \end{equation}
  where $s$ is some kinematic invariant ({\it e.g.} a Mandelstam
  variable) and $z$ is one of the integration variables. As long, as
  we are in the Euclidean regime, {\it i.e.} $s < 0$,
  Eq.~(\ref{kinematic}) contributes another oscillatory factor and
  cannot influence the convergence of the integral. For positive
  values, however, we will have

  \begin{equation}
    (-s)^{-z} = e^{-z \log(-s)} = e^{-z (\log(s) - i \pi)}
    = e^{i a \pi} \; s^z \; e^{b \pi},
  \end{equation}
  where $z = a-i b$. It is clear, that the exponential factor can
  compensate the damping from the product of gamma functions.

  An interesting example, which illustrates the problem is provided by
  the leading pole term of the first planar 7-line QED box integral,
  Fig.~\ref{B1}, which is given by

  \begin{eqnarray}
    -\frac{e^{-2 \epsilon \gamma_E}}{2
    s^2 (-t)^{1+2\epsilon}} \frac{1}{\epsilon^2} \frac{1}{(2\pi
    i)^2}&&\!\!\!\!\!\!\! \int_{-i \infty}^{i \infty}\int_{-i
    \infty}^{i \infty} dz_1\; dz_2\; 
    (-s)^{-z_1-z_2} \nonumber \\
    &\times& \frac{\Gamma^3(-z_1) 
    \Gamma(1+z_1)\Gamma^3(-z_2)\Gamma(1+z_2)}{\Gamma(-2z_1)\Gamma(-2z_2)},
  \end{eqnarray}
  where $\Re\; z_1 = \Re\; z_2 = -1/2$. This is just a product
  of two one-dimensional integrals, which can be done by closing
  contours and resumming the residues, with the result

  \begin{equation}
    \label{int1}
    \frac{1}{2\pi i}\int_{-i \infty}^{i \infty} dz_1\;(-s)^{-z_1}
    \frac{\Gamma^3(-z_1) 
    \Gamma(1+z_1)}{\Gamma(-2z_1)} =
    -\frac{4}{\sqrt{\frac{4}{s}-1}}\arcsin \sqrt\frac{s}{4}, 
  \end{equation}
  below threshold, {\it i.e.} for $0 \le s \le 4$. For $z_1 = -1/2-i
  b$, $b \gg 0$, the integrand behaves as

  \begin{equation}
    -\frac{1+i}{\sqrt{2}}\;\sqrt{\pi s}\;\frac{e^{i \log{(s/4)}\;
     b}}{\sqrt{b}}.
  \end{equation}
  As anticipated, the exponential factor disappeared. Worse even, the
  integrand is not absolutely integrable. It is interesting to note,
  that the frequency of the oscillation, $\log{(s/4)}$, encodes the
  threshold. Further examples seem to confirm that this is a general
  property. Fortunately, this integral can be evaluated using standard
  techniques for infinite range oscillatory integrands. With the
  Pantis' method \cite{kythe}

  \begin{equation}
    \int_{-\infty}^{\infty} db\; e^{-i \omega b}f(b) \simeq
    \int_{-b_0}^{\infty} db \; e^{-i \omega b}f(b) + \frac{1}{i
    \omega}e^{i \omega b_0}f(-b_0),
  \end{equation}
  setting $s = 2$ and $b_0 = 40$ the value of the integral in
  Eq.~(\ref{int1}) is $\simeq -3.17-0.09 i$, to be compared to the exact
  result, which is $-\pi$.

\begin{figure}
    \begin{center}
      \epsfig{file=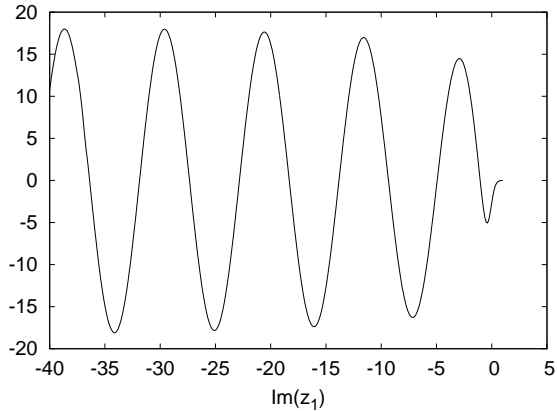, width=8cm}
    \end{center}
    \caption{\label{behav} \it Real part of the integral
    Eq.~(\ref{behaveq}) at s = 2.}
  \end{figure}

  One would be tempted to assume that the slowly convergent
  oscillatory behaviour can be factorized in one integration variable
  and that the remaining integrations are fast convergent. This
  assumption is false, as shown in Fig.~(\ref{behav}), which
  represents the integrand in $z_1$ of the original integral after
  shifting $z_1 \rightarrow z_1-z_2$ and up to normalization factors

  \begin{equation}
    \label{behaveq}
    \frac{1}{(2\pi i)^2} (-s)^{-z_1} \int_{-i \infty}^{i \infty} dz_2\;
   \frac{\Gamma^3(-z_1+z_2)
   \Gamma(1+z_1-z_2)\Gamma^3(-z_2)
   \Gamma(1+z_2)}{\Gamma(-2z_1+2z_2)\Gamma(-2z_2)}.
  \end{equation}

  Apparently, this does not seem to be integrable at all, and
  certainly no numerical method would provide a reasonable estimate,
  even if it would be integrable.

  In conclusion, one encounters massive Feynman integrals, which
  require, in the physical regime, multidimensional integration of
  slowly convergent oscillatory functions over infinite range. This
  problem can be solved, but the efficiency of the methods is very low
  and acceptable only for low dimensions. Furthermore, it might be
  necessary to shift the integration variables to obtain convergent
  representations. The latter task is certainly very difficult to
  automatize.

  The above discussion does not change anything to the fact that MB
  integrals provide reliable numerics in the Euclidean regime for all
  encountered integrals and in the Minkowski regime for the massless
  ones. It is of course not excluded that some massive integrals can
  also be done reliably without special methods, but this has to be
  checked in specific cases.

  The {\tt MB} package provides routines that can perform numerical
  integrations of MB representations. In order to work, the libraries
  {\tt libcuba.a} from {\tt CUBA} \cite{cuba}, {\tt libkernlib.a} and
  {\tt libmathlib.a} from {\tt CERNlib} \cite{cernlib} have to be
  installed either in the working directory or in a globally accessible
  directory with libraries, and the {\tt Fortran} compiler has to be
  called {\tt f77}. In case, the user wanted to change these defaults,
  it would be necessary to change the internal code of {\tt MB.m}.
 
  The main routine for numerical integration is

  \fbox{\tt MBintegrate[integrals, kinematics, options]}

  where the input arguments are

  \begin{itemize}

    \item {\tt integrals:} a list of integrals as provided by {\tt
    MBexpand}.

    \item {\tt kinematics:} a list of rules providing numeric values
    for all the parameters (usually kinematic invariants) besides the
    expansion variable and the integration variables. If the user is
    interested in Minkowski kinematics then a small imaginary part
    should be added. Even though this is just an approximation, it is
    justified by the fact, that the final result has usually much
    lower precision than the error introduced by such a procedure.

    \item {\tt options:}

      \begin{itemize}

	\item {\tt NamePrefix:} by default the Fortran programs
	  generated for integrals in more than one variable are
	  called {\tt MBpart1x0}, etc. where the last number is the power of
	  the expansion variable {\tt x} and {\tt part1} denotes the
	  first integral at this order. With this option one can
	  change the prefix {\tt MB}.

	\item {\tt PrecisionGoal, AccuracyGoal, MaxPoints,
	MaxRecursion:} numerical integration options as in
	{\tt NIntegrate}. The defaults are respectively $4$, $12$,
	$10^6$, $10^3$, and have been tuned to several problems solved
	with the package.

	\item {\tt MaxCuhreDim:} dimension threshold, 4 by default,
	  above which Vegas will be used for the evaluation of the
	  integrals instead of Cuhre.

	\item {\tt Complex:} by default, only the real part of the
	  integrals is evaluated, with this option set to {\tt True},
	  the imaginary part will also be given.

	\item {\tt FixedContours:} contours will not be shifted if
	this option is set to {\tt True}. For a detailed explanation,
	see {\tt MBshiftContours} below.

	\item {\tt NoHigherDimensional:} by default, the complete
	  integration is performed within {\tt MBintegrate}, however with
	  this option set to {\tt True}, 1-dimensional integrals are
	  evaluated and the Fortran programs are prepared, but not
	  run. This may be used to run them in parallel for example.

	\item {\tt Debug:} with this option set to {\tt True}, the Fortran
	  programs are kept after evaluation and the value of every
	  integral is given within {\tt MBval[value,  error,
	  probability, part]} objects, where {\tt value}, {\tt error} and {\tt
	  probability} are given by {\tt CUBA}, and {\tt part} is the
	  number of the integral. This provides a primitive means of
	  improving the calculation by tuning only specific
	  integrals, since the integration parameters can be easily
	  changed in the Fortran programs.

	\item {\tt Verbose:} by default the progress of the
	integration is printed to the screen. This can be switched off
	by setting this option to {\tt False}.

      \end{itemize}

  \end{itemize}

\begin{figure}
    \begin{center}
      \epsfig{file=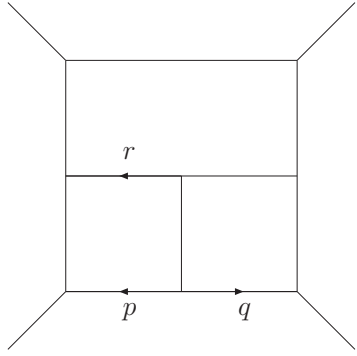, width=5cm}
    \end{center}
    \caption{\label{tennis} \it Tennis court integral, $I_4^{(3)b}$,
    containing a factor of $(p+r)^2$.}
  \end{figure}

  Instead of providing a detailed description of the output, I
  illustrate {\tt MBintegrate} on the example of the ``tennis court
  integral'', Fig.~(\ref{tennis}), introduced and calculated
  analytically in \cite{Bern:2005iz}. Since, it has never been
  confirmed independently, this example supports the correctness
  of the analytical result.

  Similarly as in \cite{Bern:2005iz}, a factor of
  $-(-s)^{-1-3\epsilon}t^{-2}$ has been taken out. If {\tt expanded}
  is the result provided by {\tt MBexpand}, which contains 65
  integrals, then the numerical evaluation proceeds as follows

  {\small
    \begin{verbatim}
In[3] := MBintegrate[expanded, {s -> -2, t -> -3}]

Shifting contours...

Performing 30 1-dimensional integrations...1...2...3...4...5...6...
7...8...9...10...11...12...13...14...15...16...17...18...19...20...
21...22...23...24...25...26...27...28...29...30

Higher-dimensional integrals

Preparing MBpart1ep0 (dim 6)

Preparing MBpart2ep0 (dim 6)
.
.
.
Preparing MBpart58ep-1 (dim 4)
.
.
.
Running MBpart1ep0

Running MBpart2ep0
.
.
.
{154.50857084232496 + 1.7777777777777777/ep^6 - 
  0.8785077342343561/ep^5 - 15.544672574293408/ep^4 - 
  20.903348302858618/ep^3 + 20.868443575404378/ep^2 + 
  84.4035478542778/ep, 
 {1.454748334713152 + 0.0012476956259284788/ep^3 + 
   0.01736836792954924/ep^2 + 0.3243732528120632/ep, 0}}
    \end{verbatim}
  }

  At a first stage, the contours are shifted with {\tt MBshiftContours},
  then the 1-dimensional integrals are evaluated in {\tt
  MATHEMATICA}. Subsequently, {\tt Fortran} programs for the higher
  dimensional integrals are prepared and run. The user can easily see
  the names of the programs and the dimensions of the
  integrals. Finally, the result is given in the form of a list. The
  first element is the result itself, whereas the second element is a
  sublist giving the errors on the real and imaginary parts respectively.
  It is important to note, that the errors are estimated from the
  square root of the sum of the squares of the errors of each of the
  higher dimensional integrals. Therefore, in the example above, the
  errors start at $1/\epsilon^3$, because up to this pole, there were
  only 1-dimensional integrals. This also implies that it is assumed
  that the error from the 1-dimensional integrals is negligible. The
  above calculation took about 1 hour on a 2.4 GHz notebook, with a 
  claimed error on the finite part of about 1\% ($1.4$ against $154.5$
  above). If compared to the exact result

  \begin{eqnarray}
    \!\!\!\! \frac{1.77778}{\epsilon^6} - \frac{0.878508}{\epsilon^5} - 
    \frac{15.5447}{\epsilon^4} - \frac{20.9033}{\epsilon^3} + 
    \frac{20.8679}{\epsilon^2} + \frac{84.4068}{\epsilon} + 154.379,
    \nonumber \\
  \end{eqnarray}
  the error is rather at the permille level. Further numerical
  evaluation examples can be found in the two notebooks provided with
  the package.

  A utility related to numerical integration, which is of interest by
  itself is

  \fbox{\tt MBshiftContours[integrals]}

  where the only argument is a list of integrals as provided by {\tt
  MBexpand}. The idea here, is that if there is a contour passing
  between two poles of a $\Gamma$ function, then the further it will
  stay from both of them, the less peaked will the integrand
  be. Since the contours have more or less random distances to the
  poles, it is wise to shift them before numerics to improve
  stability. This is achieved by the above utility.

  \subsection{Additional tools}

  \label{tools}

  Apart from performing the analytic continuation of a MB integral,
  one is usually interested in simplifying the integrals as much as
  possible. This is of utmost importance, if one is interested in
  obtaining analytic results. It is often the case, that some of the
  integrations can be performed exactly with the help of Barnes'
  lemmas.

  \underline{1st Barnes' lemma}

  \begin{eqnarray}
    \int_{-i \infty}^{i \infty} &dz&  \Gamma(a+z) \Gamma(b+z)
    \Gamma(c-z) \Gamma(d-z) =\nonumber \\ &&
    \frac{\Gamma(a+c)\Gamma(a+d)\Gamma(b+c)\Gamma(b+d)}{
    \Gamma(a+b+c+d)}.
  \end{eqnarray}

  \underline{2nd Barnes' lemma}

  \begin{eqnarray}
    \int_{-i \infty}^{i \infty} &dz& \; \frac{\Gamma(a+z) \Gamma(b+z)
    \Gamma(c+z) \Gamma(d-z) \Gamma(e-z)}{\Gamma(a+b+c+d+e+z)} =
    \nonumber \\ &&
    \frac{\Gamma(a+d)\Gamma(a+e)\Gamma(b+d)\Gamma(b+e)\Gamma(c+d)\Gamma(c+e)}{
    \Gamma(a+b+d+e)\Gamma(a+c+d+e)\Gamma(b+c+d+e)}.
  \end{eqnarray}

  Both of them are valid only if the integration contour is such that
  the poles corresponding to $\Gamma$'s with positive $z$ are
  separated from the poles with negative $z$. If this is the case, the
  user can apply the rules defined as

  \fbox{\tt barnes1[z]}\hfill and \hfill \fbox{\tt barnes2[z]},

  where {\tt z} is the integration variable. An example is the
  simplification of the integral for the $B_0$ function

  {\small
    \begin{verbatim}
In[1]:= int = b0[s, 1 + z1, 1 + z2]*ms^z1*ms^z2* Gamma[-z1]*Gamma[1 +
        z1]*Gamma[-z2]*Gamma[1 + z2] /.  z1 -> z1 - z2 /. barnes1[z2]

Out[1]:= (ms^z1*(-s)^(-ep - z1)*Gamma[1 - ep - z1]^2*Gamma[-z1]*
         Gamma[ep + z1])/Gamma[2 - 2*ep - 2*z1]
    \end{verbatim}
  }

  This is, however, a rare situation. Most of the time, the Barnes'
  lemmas are applicable to integrals after analytic continuation and
  expansion. In this case, the procedure generates integrals with
  contours parallel to the imaginary axis and the contour might not
  separate the poles of the $\Gamma$ functions. In such cases one uses
  various corollaries to the lemmas, see {\it e.g.}
  Ref.~\cite{Smirnov:1999gc}.

  \begin{figure}
    \begin{center}
      \epsfig{file=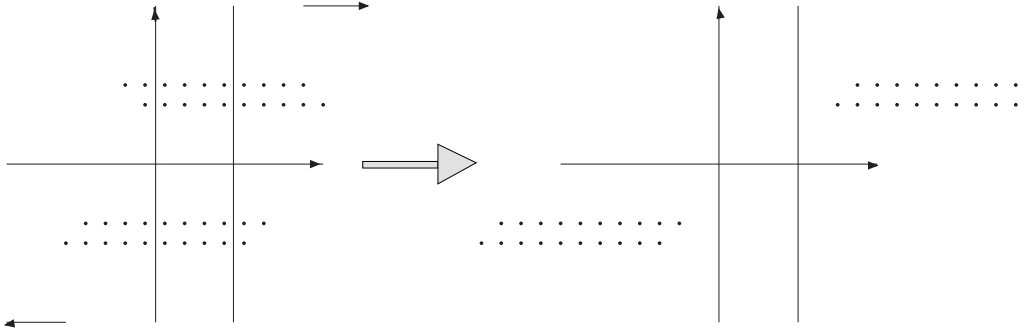, width=14cm}
    \end{center}
    \caption{\label{barnes} \it Regularization of the integrals in the
      Barnes lemmas by shifting the poles, in such a way that the contour
      separates the poles.}
  \end{figure}
  
  I propose here an automatic procedure based on {\tt MBcontinue}. The
  idea is to shift all of the $a,b, ...$ variables by $\epsilon$, such
  that the contours be separated and then analytically continue with
  $\epsilon$ to $0$, as illustrated in Fig.~\ref{barnes}. In the case
  of the first Barnes lemma, the shift is determined by the condition

  \begin{equation}
    \epsilon > \max\left(-\min(a,b)-\Re(z),\;\Re(z)-\min(c,d)\right).
  \end{equation}

  This algorithm is implemented in the following two routines:

  \fbox{\tt Barnes1[MBint[integrand,\{fixedVarRules,intVarRules\}],z]}
  
  and

  \fbox{\tt Barnes2[MBint[integrand,\{fixedVarRules,intVarRules\}],z]}
  
  The arguments are as described in {\tt MBcontinue} and {\tt z} is
  the integration variable that should be eliminated by Barnes'
  lemma. An example usage can be found in {\tt MBexamples2.nb}.

  In the case, where the integral contains a $\psi$ function, the user
  has to apply the lemma to the corresponding integrand with a
  $\Gamma$ function and only then derive the result. This might be
  automatized in the future.

  Of lesser importance are the remaining tools. To help in the
  construction of efficient MB integrals, there are several known
  exact expressions for the $A_0$, $B_0$ and $C_0$ functions taken
  from the Appendix of Ref.~\cite{Smirnov:2004ym}. Details can be
  found directly in the code of {\tt MB.m}. In case the user would
  like to construct his representation directly from a Feynman
  parameter integral as is done {\it e.g.} in
  Ref.~\cite{Tausk:1999vh}, there is also a routine

  \fbox{\tt FUPolynomials[integrand, momenta, invariants]}

  that generates the $F$ and $U$ polynomials in the notation of
  Ref.~\cite{Binoth:2000ps}. The input is
  
  \begin{itemize}

    \item {\tt integrand:} a product of propagators {\tt DS[k,m,n] =}
    $1/(k^2-m^2)^n$.

    \item {\tt momenta:} the loop momenta.

    \item {\tt invariants:} a list of rules, {\it e.g.} {\tt p1*p2 ->
    1/2*s-m\^{}2}, which transform products of external momenta into
    some suitable notation, for example the Mandelstam variables.

  \end{itemize}

  In the output, one obtains a list of four elements. First come the
  $F$ and $U$ polynomials, then the $M$ matrix and $Q$ vector again in
  the notation of Ref.~\cite{Binoth:2000ps}.

  \section{Conclusions}

  I presented a practical tool for automatic analytic continuation of
  MB integrals. It can be used either as part of a Feynman diagram
  calculation leading to an analytic result in terms of some known
  functions, or as a tool for directly providing numerical
  results. Irrespective of the aim, the most cumbersome part of the MB
  technique has been reduced to a mere use of one {\tt MATHEMATICA}
  function, making high order calculations in perturbation theory
  significantly easier and more accessible to the interested.

  \begin{ack}

    I would like to thank J. Gluza for testing the package and
    V. A. Smirnov for motivating me to make it public through the
    present work. The development of this package profited very much
    from a long collaboration with J. Gluza and T. Riemann on the NNLO
    corrections to Bhabha scattering in QED.

    This work was supported by the Sofja Kovalevskaja Award of the
    Alexander von Humboldt Foundation sponsored by the German Federal
    Ministry of Education and Research, and by the Polish State
    Committee for Scientific Research (KBN) for the research project
    in years 2004-2005.

  \end{ack}

\end{document}